\documentclass[12pt,reqno]{amsart}
\usepackage{graphicx}
\usepackage{setspace}
\usepackage{empheq}
\usepackage{cite}
\usepackage{amsmath}
\usepackage{dutchcal}
\usepackage{textcomp}
\usepackage{amssymb}

\usepackage{color}
\definecolor{MyLinkColor}{rgb}{0,0,0.4}

\newcommand{\e}{\varepsilon}

\newcommand{\f}{\frac}

\theoremstyle{remark}

\numberwithin{equation}{section}

\begin{document}



\begin{center}

\section*{\LARGE \bf{On the modelling of short and intermediate water waves }}

\vskip1cm

{\large \bf  Rossen I. Ivanov \footnote{member of the Institute for Advanced Physical Studies,
111 Tsarigradsko shose Blvd., Sofia 1784, Bulgaria }}
\vskip1cm

\hskip-.3cm
\begin{tabular}{c}
\\
{\it School of Mathematics and Statistics,Technological University Dublin, }
\\
{\it City Campus, Grangegorman, Dublin D07 ADY7, Ireland}
\\
\\{\it e-mail: rossen.ivanov@tudublin.ie  }
\\
\end{tabular}
\end{center}

\vskip0.4cm

\input epsf

{\small
{\bf Abstract:} { The propagation of water waves of finite depth and flat bottom is studied in the case when the depth is not  small in comparison to the wavelength. This propagation regime is complementary to the long-wave regime described by the famous KdV equation.    
 The Hamiltonian approach is employed in the derivation of a model equation in evolutionary form, which is both nonlinear and nonlocal, and most likely not integrable. Possible implications for the numerical solutions are discussed.}
\vskip0.3cm
{\bf Mathematics Subject Classification (2010):} 76B15, 35Q35, 37K05
\vskip0.3cm
{\bf Keywords:} nonlinear waves, short waves, intermediate waves, nonlocal differential equations
}

\section{Introduction}

Most of the model equations for water waves are approximations for the long-wave propagation regimes, since most of the energy of the wave motion is concentrated in these waves. Long waves (or shallow-water waves) are defined usually as the depth to wavelength ratio $\delta= h/\lambda <0.05.$ Several famous integrable nonlinear equations, like the KdV equation \cite{KdV,ZMNP}, are models for long waves of small amplitude. The short waves (or waves over deep water) are usually defined with $\delta>0.5,$ and the intermediate waves (or transitional waves) - with $0.05<\delta<0.5.$ The intermediate and short waves received a lot less attention, and one reason is perhaps the fact that the corresponding approximations lead to more complicated, nonlinear and nonlocal equations. In \cite{BS71} an integral equation for surface waves has been proposed for arbitrary wavelengths and finite depth. The problem has been studied in \cite{Mats} and model equations both for long and short waves are derived from the governing equations as well. 
The short-wave effects usually compete with the capillarity effects and then resonances can be observed - these have been studied quite a lot, see for example \cite{Phil,MG,Kart,CK,MarJmfm,CaCh,Cra,IM}.

For the intermediate long waves or for waves on deep water the so-called Benjamin-Ono (BO) \cite{BO1,BO2, CoIv} and the Intermediate Long Wave Equation (ILWE) \cite{J,K,CuIv} are derived for the internal waves below a flat surface, which leads to some simplifications and these models are in fact integrable. 

The aim of this work is to illustrate one application of the Hamiltonian approach in the derivation of water waves model for a single layer of water in the case of small amplitude short and intermediate waves. 

Following the seminal paper of Zakharov \cite{Zak}, the Hamiltonian approach for water waves propagation has been developed extensively, see for example \cite{Broer,Mil1,Mil2, BO,Rad,NearlyHamiltonian}. A convenient explicit representation of the Hamiltonian involves the non-local Dirichlet-Neumann operator,  e.g. \cite{Craig1993,CraigGroves1,CGK}.  In our derivation we are using the Hamiltonian approach for a single layer gravity waves, and only the assumption for a small amplitude compared to depth has been made.

\section{Preliminaries}
\subsection{The governing equations}

\begin{figure}[!ht]
\centering
\includegraphics[width=0.8 \textwidth]{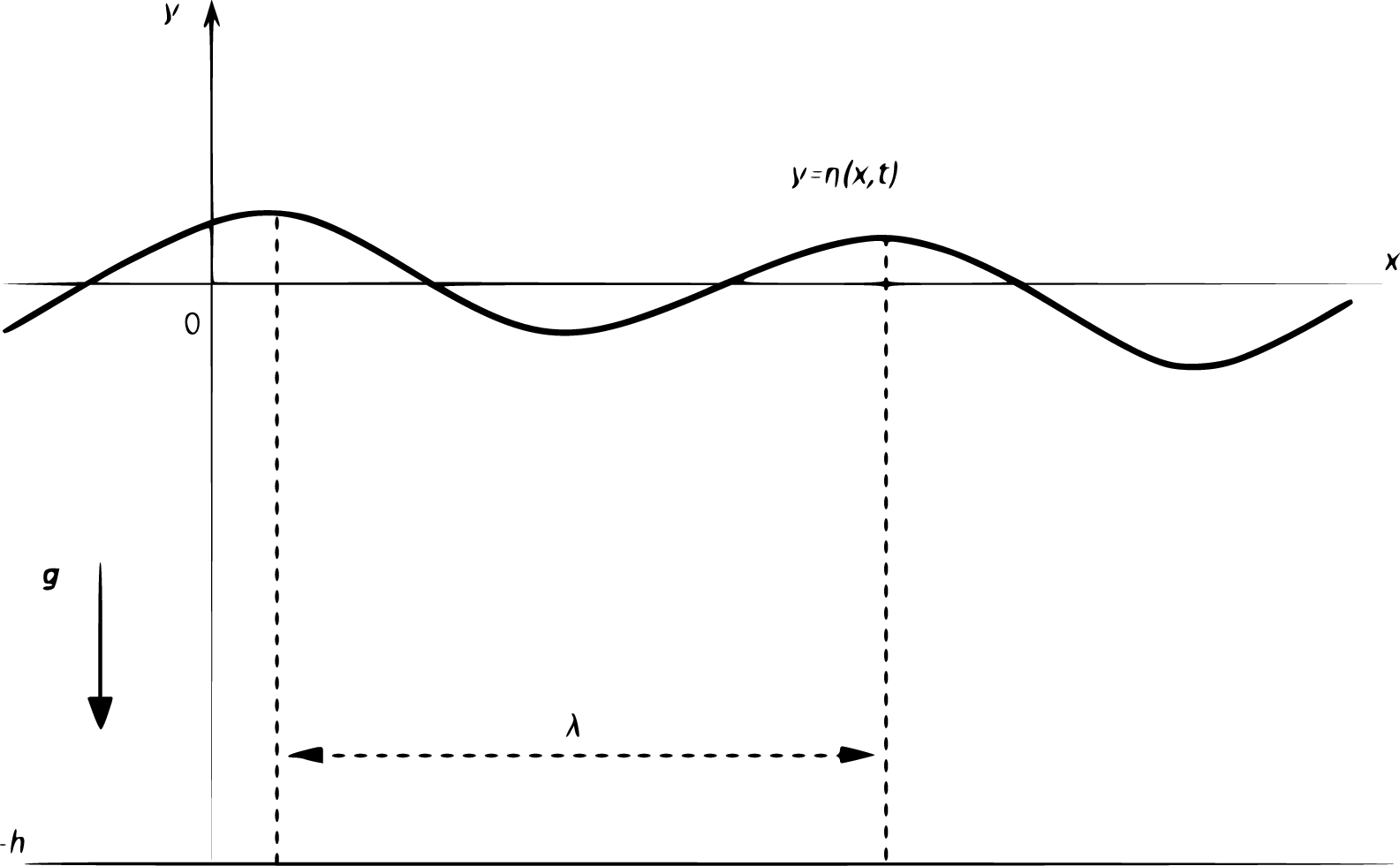}
 \caption{Coordinates and general setup.}\label{fig1}
\end{figure}

We choose a Cartesian coordinate system with a horizontal coordinate $x$ and vertical coordinate $y.$
We denote with $t$ the time variable. We recall briefly the governing equations for gravity water waves with a free surface (denoted $y=\eta(x,t)$) of a two dimensional irrotational water flow bounded below by a flat bed $y=-h$ ($h$ being some positive constant) and above by the surface itself, see Fig. \ref{fig1}.
Denoting with $(u(x,y,t),v(x,y,t))$ the velocity field, with $P(x,y,t)$ the pressure and with $g$ the gravitational constant the equations of motion are the Euler equations (unit density)
\begin{equation}\label{Eulereq}
\begin{split}
u_t+uu_x+vu_y&=-P_x\\
v_t+uv_x+vv_y&=-P_y-g,
\end{split}
\end{equation}
and the equation of mass conservation
\begin{equation}\label{masscons}
u_x+v_y=0.
\end{equation}
The systems \eqref{Eulereq} and \eqref{masscons} are complemented by the kinematic boundary conditions
\begin{equation}\label{BKS}
\begin{split}
v&=\eta_t + u \eta_x\quad{\rm on}\quad y=\eta(x,t),\\
v&=0\quad{\rm on}\quad y=-h,
\end{split}
\end{equation}
and $P=P_{\rm atm} $ on $ y=\eta(x,t).$ The fluid motion in the absence of vorticity can be expressed through the velocity potential $\varphi$ as
\begin{equation}
u=\varphi_x(x,y,t), \qquad v=\varphi_y(x,y,t),
\end{equation} The zero vorticity within the flow is chosen for simplicity, that is $u_y-v_x=0,$ the surface tension is neglected as well.

Introducing $\xi(x,t):=\varphi(x,\eta(x,t),t)$, the one-dimensional wave propagation on the surface in the $x$ direction can be expressed in a Hamiltonian form  \cite{Zak,BO,CraigGroves1}
\begin{equation}\label{NH}
 \eta_t = \f{\delta H}{\delta\xi} \quad {\rm and}\quad \xi_t= -\f{\delta H}{\delta\eta}.
\end{equation} where $H[\xi,\eta]=\tilde{H}/\rho$, where $\rho$ is the fluid density and 
\begin{equation}
\tilde{H}= \frac{1}{2}\rho \int_{\mathbb{R}}  \int_{-h}^{\eta}(u^2+v^2)dy dx+\rho g   \int_{\mathbb{R}}  \int_{-h}^{\eta}y dydx.
\end{equation}
is the total energy of the fluid, which could be written in terms of the surface variables $\xi, \eta.$

\subsection{The Hamiltonian for the short and intermediate gravity water waves}

In the case of gravity waves the Hamiltonian could be expressed through the surface variables as
\begin{equation}
H[\xi, \eta]=\frac{1}{2}\int_{\mathbb{R}} \xi G(\eta)\xi\,dx+\frac{g}{2}\int _{\mathbb{R}} \eta^2\,dx
\end{equation} where $G(\eta)$ is a nonlocal self-adjoint operator, known as Dirichlet-Neumann operator, see more details for example in \cite{CraigGroves1}. The Hamiltonian variables $\xi,$ $\eta$  are assumed Schwartz class functions with respect to both arguments, so that the integrals and operators under consideration make sense.

We introduce, as usual, the scale parameters $\varepsilon =a/h$ and $ \delta = h/\lambda$ where $a$ is the wave amplitude ($0<|\eta(x,t)|\le a$) and $\lambda$ is the wavelength. For small amplitude short and intermediate wavelengths the scaling is $\varepsilon \ll 1$ but $\delta$ can not be considered small, and the Dirichlet-Neumann operator has a perturbative expansion in the form
$$G=G_0+\varepsilon G_1+\varepsilon^2 G_2+\ldots.$$ In what follows we need only the first two terms,
\begin{equation}
    G(\eta)=D\tanh(hD)+ \varepsilon(D\eta D -D\tanh(hD) \eta D\tanh(hD))+ \mathcal{O}(\varepsilon ^2),
\end{equation}
where $D=-i\partial_x\equiv -i \partial.$ Let us introduce for convenience  $\mathcal{T}:=-i \tanh(hD)$, then 
\begin{equation}
    G(\eta)=\partial \mathcal{T} - \varepsilon \partial \eta \partial -\varepsilon \partial \mathcal{T} \eta \mathcal{T} \partial+ \mathcal{O}(\varepsilon ^2),
\end{equation}

The other scales are as follows. The wave elevation, $\eta \simeq \varepsilon h$ is of order $\varepsilon$, and $\xi$ is of the same order as $\eta$, i.e. of order $\varepsilon$. The operator $hD$ has a Fourier multiplier $hk\simeq 2 \pi h/ \lambda \simeq 2 \pi $  as far as $\delta \simeq 1$ then $ \partial_x$ is of the order of $ (1/h) (hD)\simeq 2\pi/h$ which is a fixed constant and therefore is $ \mathcal{O}(1) .$ The time derivative is assumed to be also of order $1$. Therefore, writing explicitly the scale parameter, and keeping terms up to $\varepsilon^3$ we obtain
\begin{equation}\label{Ham}
 H(\eta,\xi)=\f{\varepsilon^2}{2}\int \xi \mathcal{T}\xi_x\,dx+\f{\varepsilon^2 g}{2}\int\eta^2 \,dx+\f{\e^3}{2}\int \eta\xi_x^2\,dx-\f{\e^3}{2}\int \eta (\mathcal{T}\xi_x)^2\, dx.
\end{equation}










\section{The main result}

\subsection{Nonlinear Hamiltonian equations }

From \eqref{NH} and \eqref{Ham} we obtain the following system of coupled equations
\begin{align*}
    \xi_t&=-\left(g\eta+\frac{\varepsilon}{2}\xi_x^2-\frac{\varepsilon}{2}(\mathcal{T}\xi_x)^2 \right),\\
    \eta_t&=\mathcal{T}\xi_x-\varepsilon (\eta\xi_x)_x - \varepsilon \mathcal{T}(\eta \mathcal{T}\xi_x)_x.
\end{align*} or, written through the variable $\mathfrak{u}=\xi_x,$ 
\begin{align}
   &\mathfrak{u}_t +g\eta_x+\frac{\varepsilon}{2}\left(\mathfrak{u}^2  -(\mathcal{T}\mathfrak{u})^2\right)_x =0,\\
    &\eta_t-\mathcal{T}\mathfrak{u}+\varepsilon (\eta\mathfrak{u})_x + \varepsilon \mathcal{T}(\eta \mathcal{T}\mathfrak{u})_x=0. \label{eta2}
\end{align} 
Now we introduce a new variable, $\mathfrak{v}=\mathfrak{u}-\varepsilon \eta_x \mathcal{T}\mathfrak{u}_x.$ Noticing that $\mathfrak{v}=\mathfrak{u}+\mathcal{O}(\varepsilon),$ and neglecting terms of order $\varepsilon^2$ we obtain the system 
\begin{align}
   &\mathfrak{v}_t +g\eta_x+\varepsilon\mathfrak{v}\mathfrak{v}_x  -\varepsilon g \eta_x \mathcal{T}\eta _x =0,\\
    &\eta_t-\mathcal{T}\mathfrak{v}+\varepsilon (\eta\mathfrak{v})_x + \varepsilon \mathcal{T}(\eta \mathcal{T}\mathfrak{v})_x=0. \label{etat}
\end{align}
Next, we differentiate \eqref{etat} with respect to $t.$ Using the asymptotic expansions $\mathfrak{v}_t=-g\eta_x + \mathcal{O}(\varepsilon),$ $\eta_t=\mathcal{T}\mathfrak{v}+\mathcal{O}(\varepsilon)$ and 
$\mathcal{T}\mathfrak{v}=\eta_t+\mathcal{O}(\varepsilon)$ where possible, 
and neglecting terms of order $\varepsilon^2$ we obtain a single equation for $\eta,$ 
\begin{equation}
    \eta_{tt}+g\mathcal{T}\eta_x-\varepsilon\left[ g \eta \eta_x + g \mathcal{T}(\eta \mathcal{T}\eta_x) 
    -\eta_t \mathcal{T}^{-1} \eta_t - \frac{1}{2}\mathcal{T}(\mathcal{T}^{-1}\eta_t)^2-\frac{1}{2} \mathcal{T}(\eta_t^2)\right] _x=0.
\end{equation}
This equation describes the surface motion of all types of waves, including the short and intermediate waves. As it could be seen this is a complicated nonlinear and nonlocal equation since $\mathcal{T}$ in principle involves infinitely many derivatives. 

A further simplification comes in the case of short waves (or waves over deep water), when $\delta=h/\lambda \ge 0.5,$ then 
$|\tanh(kh)|\ge \tanh(\pi)= 0.99627, $  thus for the given range of $\delta ,$  we observe that $|\tanh(kh)|$ approaches 1. Therefore we use the approximation $\mathcal{T}=-i\tanh(hD) \approx -i\text{sgn}(D)\equiv \mathcal{H},$  where $\mathcal{H}$ is the well known Hilbert transform, defined by
\begin{equation} \label{HT}
\mathcal{H}\{f\} (x) := \mathrm{P.V.}\frac{1}{\pi}\int_{-\infty}^{\infty}\frac{f(x')dx'}{x-x'}.
 \end{equation}
It has also the property $\mathcal{H}^{-1}=-\mathcal{H},$ therefore the equation in this case becomes
\begin{equation}\label{3.7}
    \eta_{tt}+g\mathcal{H}\eta_x-\varepsilon\left[ g \eta \eta_x + g \mathcal{H}(\eta \mathcal{H}\eta_x) +\eta_t \mathcal{H} \eta_t - \frac{1}{2}\mathcal{H}(\mathcal{H}\eta_t)^2-\frac{1}{2} \mathcal{H}(\eta_t^2)\right] _x=0.
\end{equation}
Using the following identity about the Hilbert transforms 
$$\mathcal{H}(f^2)=\mathcal{H}(\mathcal{H}f)^2 +2f\mathcal{H}f$$ we can rewrite the equation \eqref{3.7} in the form
\begin{equation}\label{etatt}
   \eta_{tt}+g\mathcal{H}\eta_x-\varepsilon\left[ g \eta \eta_x + g \mathcal{H}(\eta \mathcal{H}\eta_x) -\mathcal{H}(\mathcal{H}\eta_t)^2 \right] _x=0.  
\end{equation}

This is the equation obtained in \cite{Mats} by other methods from the governing equations. The equation is second order in $t$, and in addition, it is nonlinear and nonlocal, because of the nature of the Hilbert transform.

\subsection{Evolution equation}

The author of \cite{Mats}, Prof. Matsuno has posed the problem whether \eqref{etatt} could be written in an evolutionary form, like for example the nonlinear KdV equation \cite{KdV}. The evolutionary form which involves one time derivative is definitely advantageous for numerical solutions in comparisson to the second-order in time \eqref{etatt}. Here we show that this is possible, however the equation still remains highly nonlocal.

The first observation is that the dispersion relation is $c^2(k)=g/|k|,$ or  $c(k)=\pm \sqrt{g/|k|}.$  The two signs are for the left- and the right-running waves. Let us take the plus sign and analyse the right-running waves. In operator form the wave-speed is $\hat{c}=\sqrt{g}|D|^{-1/2},$ where $|D|=D \text{sgn}(D)=-i\text{sgn}(D)\partial=\mathcal{H}\partial.$
Thus we expect to obtain an evolutionary equation in the form 
\begin{equation}\label{evolAnz}
    \eta_t + \hat{c}\eta_x+\varepsilon \mathcal{F}[\eta]=0
\end{equation}
where $\mathcal{F}$ is some expression defined via $\eta.$ Using the fact that $\partial_t=-\hat{c}\partial_x+\mathcal{O}(\varepsilon),$ we differentiate \eqref{evolAnz} with respect to $t$ and obtain
\begin{align}
    & \eta_{tt}+\hat{c}\eta_{xt}+\varepsilon \mathcal{F}_t=0, \\
    &\eta_{tt}+\hat{c}(-\hat{c}\eta_{x}-\varepsilon \mathcal{F}    )_x -\varepsilon 
    \hat{c}\mathcal{F}_x = \mathcal{O}(\varepsilon^2),   \\
    &\eta_{tt}-\hat{c}^2\eta_{xx}-2\varepsilon (\hat{c} \mathcal{F})_x=\mathcal{O}(\varepsilon^2).
\end{align}
The comparison with \eqref{etatt} yields
\begin{equation}
    \mathcal{F}[\eta]=\frac{1}{2}\hat{c}^{-1}\left[ g \eta \eta_x + g \mathcal{H}(\eta \mathcal{H}\eta_x) -\mathcal{H}(\mathcal{H} \hat{c}\eta_x)^2 \right].
\end{equation}

Thus we obtain a highly nonlocal equation in an evolutionary form:

\begin{equation}\label{EE}
     \eta_t + \hat{c}\eta_x+ \frac{\varepsilon}{2}\hat{c}^{-1}\left[ g \eta \eta_x + g \mathcal{H}(\eta \mathcal{H}\eta_x) -\mathcal{H}(\mathcal{H} \hat{c}\eta_x)^2 \right]=0.
\end{equation}

The equation is (most likely) not integrable, and in contrast to the famous integrable KdV or NLS equations, it looks rather complicated - as a matter of fact it is strictly speaking an integro-differential equation. The possible numeric solution would require a Fourier transform at every time step, then calculation of the corresponding Fourier multipliers and eventually evaluation of all values of the function at the next time step. Introducing a Fourier transform,
\begin{equation}
    \eta(x)=\frac{1}{2\pi} \int e^{ikx} E(k) dk, \qquad E(k)=\int e^{-ikx} \eta(x) dx,
\end{equation}
after a substitution in \eqref{EE} we obtain an integro-differential equation for $E(k,t):$

\begin{equation}
    -iE_t+\text{sgn}(k) \sqrt{g|k|}E+\frac{\varepsilon\sqrt{g|k|}}{4\pi}\int \left[q+\text{sgn}(k)(\sqrt{|q||k-q|}-|q|)\right ] E(k-q)E(q)\, dq.
\end{equation}
This form of the equation may have advantages for the application of numerical methods of solutions.

\subsection{Example }

Since the model equation \eqref{EE} is nonlocal, it is not possible to formulate a proper initial value problem. What is possible is to generate a solution working perturbatively, starting from some solution $\eta^{(0)}(x,t)$ of the linearised equation. Let us take for simplicity $\eta^{(0)}(x,t)=\eta_0 \cos(k_0 x - \sqrt{g|k_0|} t)$ which solves the equation in its leading order (linear approximation), $\eta_0,k_0$ are constants. Then 
$$E^{(0)}=\pi\eta_0(\delta(k-k_0)+\delta(k+k_0))e^{-i \text{sgn}(k)\sqrt{g|k|}t}.$$
looking for a solution 
\begin{align}
& \eta(x,t)=\eta^{(0)}(x,t)+\varepsilon \eta^{(1)}(x,t)+\ldots, \\
& E(k,t)=E^{(0)}(k,t)+\varepsilon E^{(1)}(k,t)+\ldots
\end{align}
we obtain 
\begin{align}
& E^{(1)}(k,t)=\frac{\pi \eta_0^2 k_0}{4(\sqrt{2}-1)}\left(  \delta(k-2k_0)+\delta(k+2k_0)\right)e^{-2i \text{sgn}(k)\sqrt{g|k_0|}t},\\
& \eta^{(1)}(x,t)=\frac{\eta_0^2 k_0 }{4(\sqrt{2}-1)} \cos(2 k_0 x - 2\sqrt{g|k_0|} t).
\end{align}
This is the second harmonic, and its amplitude is smaller by a factor of $\varepsilon.$ Apparently the nonlinearities generate (in principle) all other multiple modes, but the smallness of their amplitudes makes them insignificant.

\section{Discussion}
We point out that other quantities could be determined from the solution $\eta(x,t).$   
From \eqref{EE} and \eqref{eta2} by excluding $\eta_t$ one can find an expression for the ''tangential'' velocity $\mathfrak{u}$ in terms of $\eta:$
\begin{equation}
    \mathfrak{u}=\hat{c}\mathcal{H}\eta_x-\varepsilon \mathcal{H}(\eta \hat{c}\mathcal{H}\eta_x)_x-\varepsilon \mathcal{H} \mathcal{F}[\eta]+\mathcal{O}(\varepsilon ^2)
\end{equation}

Furthermore, it is known that in the irrotational case from the surface variables $(\eta,  \mathfrak{u})$ one can recover the variables like the pressure and the velocity potential $\varphi(x,y)$ in the bulk of the fluid.

The solitary waves are stable travelling wave solutions $\eta(x-\mathfrak{c}t),$ which are characterised by a constant speed $\mathfrak{c}$ and fast decay of the wave profile, $\eta(x-\mathfrak{c}t)\to 0$ at $x\to\pm \infty.$ Equation \eqref{etatt} in this case simplifies to
\begin{equation} \label{trwave}
    \mathcal{H }\eta ' =\frac{g}{\mathfrak{c}^2}\left(\eta+ \varepsilon \mathcal{H}(\eta \eta')\right) +\mathcal{O}(\varepsilon ^2)
\end{equation}
The solitary waves have a different nature in comparison to the periodic waves. Their existence has been studied in \cite{vera} where the claim is that solitary waves for deep water do not exist.
The result is based on the analysis of an equation, which is asymptotically equivalent to \eqref{trwave}. A particular parameterisation of the water surface is used, which introduces a new variable, see also \cite{Buff}.
The presence of surface tension changes the situation and capillary-gravity waves of solitary type on
deep water do exist. This has been proven in \cite{Ioos}, although numerically these have been studied previously for example in \cite{LH,VdB}.

Additional effects like vorticity, currents and surface tension can also be handled in the Hamiltonian framework and included as generalisations, see for example in \cite{NearlyHamiltonian,IM} and this will be accomplished in a separate publication.

\subsection*{Acknowledgements} The author is thankful to two anonymous referees for their valuable comments, feedback and advice. The author acknowledges the support from grant 21/FFP-A/9150, Science Foundation Ireland.

\end{document}